\documentclass[a4paper,twocolumn,prl,showpacs]{revtex4}
\usepackage{graphicx}

\renewcommand{\[}{\begin{equation}}
\renewcommand{\]}{\end{equation}}
\def\bea{\begin{eqnarray}}
\def\eea{\end{eqnarray}}
\def\nn{\nonumber\\}
\newcommand{\equ}[1]{Eq.~(\ref{#1})}
\newcommand{\eqs}[2]{Eqs.~(\ref{#1}) and (\ref{#2})}

\def\ev#1{\langle#1\rangle}
\renewcommand{\P}{{\bf P} }
\newcommand{\E}{{\bf E}}

\renewcommand{\r}{{\bf r}}
\newcommand{\R}{{\bf R}}
\def\prb#1#2#3{Phys. Rev. B {\bf #1}, #2 (19#3)}

\begin{document}

\title{Towards a bulk theory of  flexoelectricity}

\author{Raffaele Resta}

\affiliation{Dipartimento di Fisica, Universit\`a di Trieste, Italy, \\
and DEMOCRITOS National Simulation Center, IOM-CNR, Trieste, Italy}

\date{}

\begin{abstract}
Flexoelectricity is the linear response of polarization to a strain gradient. Here we address the simplest class of dielectrics, namely elemental cubic crystals, and we prove that therein there is no extrinsic (i.e. surface)
contribution to flexoelectricity in the thermodynamic limit. The flexoelectric tensor is expressed as a bulk response of the solid, manifestly independent of surface configurations. Furthermore, we prove that the flexoelectric responses induced by a long-wavelength phonon and by a uniform strain gradient are identical.
\end{abstract}


\maketitle 

Flexoelectricity is by definition the linear response of polarization $\P$ to strain gradient; it is measured by the fourth-rank Cartesian tensor $\mu_{\alpha\beta\gamma\delta}$, i.e. \[ P_\alpha=\mu_{\alpha\beta\gamma\delta} \, \frac{\partial \epsilon_{\gamma\delta}}{\partial r_\beta} , \label{def} \] where summation over repeated indices is understood; such tensor property is symmetry allowed in any dielectric. In recent years, there has been much interest in creating piezoelectric composites from materials which are not themselves piezoelectric, by exploiting flexoelectricity~\cite{Zhu06}. There have been also some measurements~\cite{Zubko07}, atomistic calculations~\cite{Maranganti09}, and even (very recently) first-principle calculations for BaTiO$_3$ and SrTiO$_3$~\cite{Hong10}.
However, the basic issue whether flexoelectricity is a pure bulk effect---or instead it has a contribution which depends on the surface conditions of the sample---is unsettled to date, even for the simplest of crystalline dielectrics. It is worth mentioning that the analogous issue about the simpler case of piezoelectricity is far for being trivial either~\cite{Martin72}, and spurred much discussion among theorists until two decades ago~\cite{debate}. The issue whether a material property is a bulk effect is of the utmost importance for theorists, since in the affirmative it can be in principle computed within periodic boundary conditions: for piezoelectricity, this is happening routinely since 1989~\cite{rap54}. The most quoted theoretical paper on flexoelectricity, by Tagantsev~\cite{Tagantsev86}, was published in 1986. Tagantsev's message is opposite to the present one, even in the abstract; a brief critique of Tagantsev's work is provided at the end of this work.

The present work is a first step towards a bulk theory of flexoelectricity. Here we limit ourselves to the simplest possible case: elemental cubic crystals, having a primitive lattice: we prove that flexoelectricity is indeed a genuine bulk property therein: there is no surface contribution to the flexoelectric response. The class of crystals addressed here includes solid rare gases and little more; therefore the present result looks having only academic interest. However, we believe that the additional effects (due to sublattice displacements and ionic screening) which occur in nonprimitive lattices can be also tamed at the price of a much clumsier algebra. This is briefly discussed towards the end of this work.

The definition of \equ{def} is incomplete, since one has to specify the
macroscopic field $\E$.  It is customary to {\it define} the material constants (such as Born effective charges, piezoelectric and pyroelectric constants) as derivatives of $\P$ with respect to the relevant variable,  taken at {\it zero} field. If we make such a choice in \equ{def}, then the flexoelectric polarization in a nonzero field $\E$ is then \[ P_\alpha=\mu_{\alpha\beta\gamma\delta} \, \frac{\partial \epsilon_{\gamma\delta}}{\partial r_\beta}  + \frac{\varepsilon_\infty -1}{4\pi} E_\alpha , \label{def2} \] where we have exploited cubic symmetry and 
$\varepsilon_\infty$  is the dielectric constant ($\varepsilon_0 = \varepsilon_\infty$ in primitive lattices). Whenever $\mu$ is a bulk effect, then  \equ{def2} does not depend on the shape of the sample; the only effect of the shape is represented by a depolarization field and included in $\E$~\cite{Landau,rap_a30}. For a given shape (free-standing in zero external field), the depolarization field dictates the surface charge at the boundary, via Gauss theorem. Conversely, when a genuine surface effect contributes to the apparent polarization, some extra charge must accumulate at the boundary. Here we choose the worse possible shape (see below), where the depolarization field is maximum, and we show that even in this case there is no extra boundary charge. The absence of surface effects is confirmed by the case (discussed below) of a crystal with no boundary and a built-in long-wavelength phonon.

\begin{figure}
\centerline{\includegraphics[width=.5\textwidth]{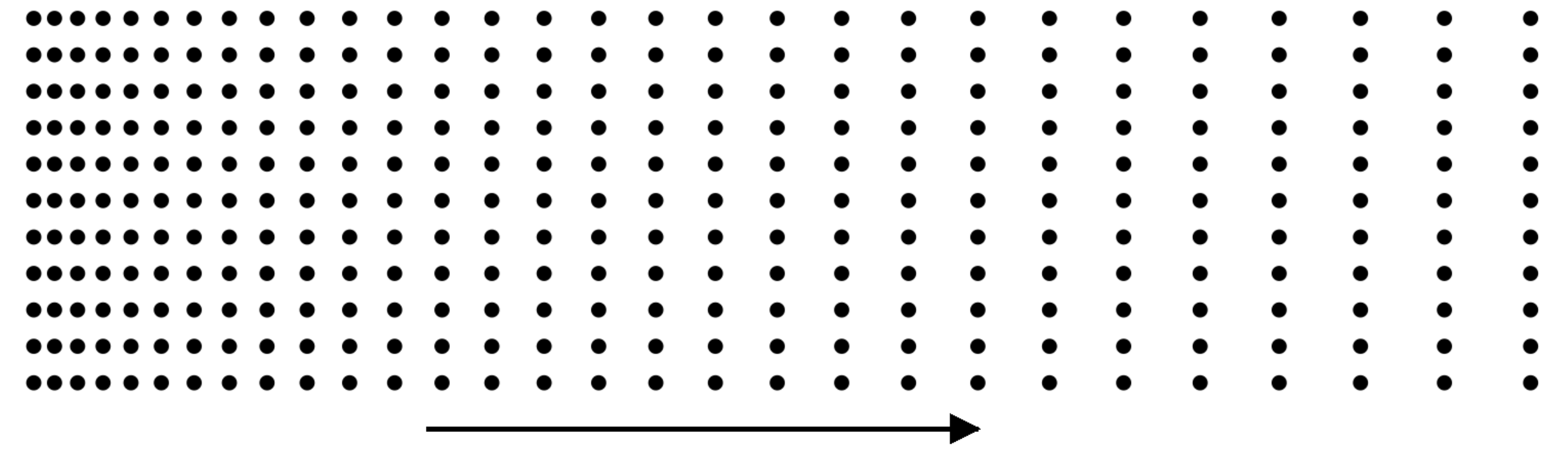}}
\bigskip
\centerline{\includegraphics[width=.5\textwidth]{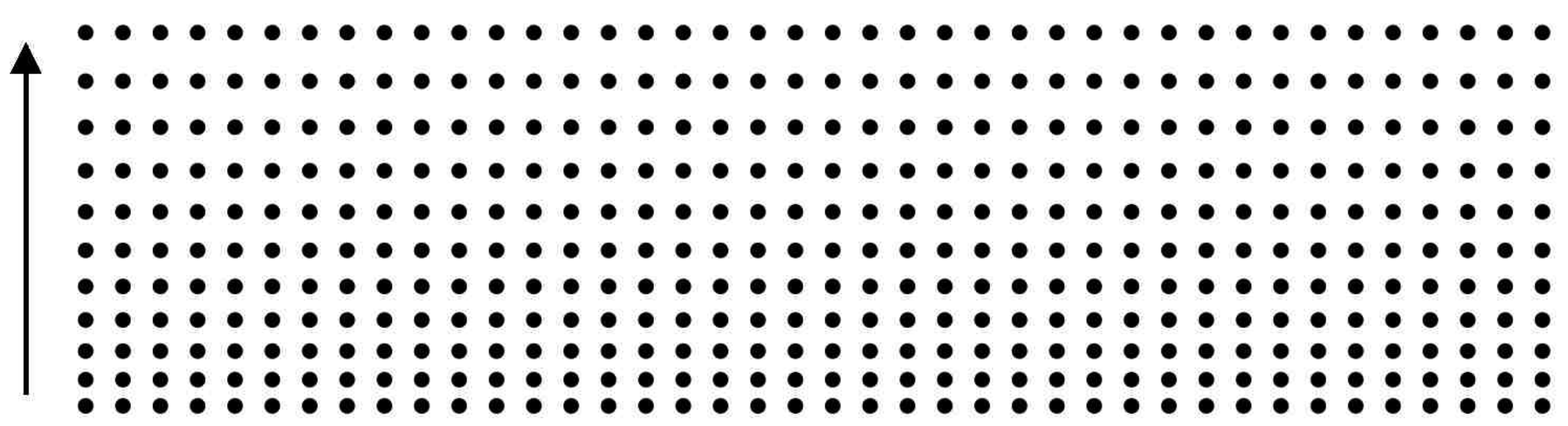}}
\caption{Slabs with a built-in uniform strain gradient (uniaxial); the $x$-direction is indicated by the arrow in each panel. Top: ``transverse'' case. Strain, strain gradient, and polarization are parallel to the slab, while the field vanishes. Bottom: ```longitudinal'' case. Strain, strain gradient, polarization, and depolarizing field are all normal to the slab. Similar figures can be drawn even when $x$ is not along a cubic axis.}
\label{fig:slabs} \end{figure}

In a crystal of cubic symmetry the tensor $\mu_{\alpha\beta\gamma\delta}$ depends on three independent elements: these can be evaluated in principle by considering the uniaxial response---where strain, strain gradient, and polarization are all aligned---along three different directions, e.g. (100), (110), and (111). For the sake of simplicity, in the following we label as ``$x$'' any of these directions, 
and we address the uniaxial flexoelectric coefficients, i.e.
\[ P_x=\mu_{xxxx} \, \frac{\partial \epsilon_{xx}}{\partial x}  + \frac{\varepsilon_\infty -1}{4\pi} E_x  \label{uni} . \] 
We consider a sample in the form of a slab, where $\partial \epsilon_{xx}/\partial x$ is either parallel or normal to the slab: this is schematically shown in Fig.~\ref{fig:slabs}. In the former case $\E$ vanishes and $\P$ is parallel to $x$ (``transverse''); in the latter case both $\E$ and $\P$ are parallel to $x$, and are related by $\E = - 4\pi \P$ (``longitudinal'')~\cite{Landau,rap_a30}. We stress that such properties are a consequence of cubic bulk symmetry, even if $x$ is not a cubic direction (Fig.~\ref{fig:slabs} only shows cubic directions for the sake of simplicity). Eventually, we will address infinitely thick slabs: the two thermodynamic limits {\it are different} because of the long range of Coulomb interaction.

We focus on the longitudinal case: replacement of $\E = - 4\pi \P$ into \equ{uni} yields \[ E_x = - \frac{4 \pi \mu_{xxxx} }{\varepsilon_\infty} \frac{\partial \epsilon_{xx}}{\partial x} . \label{reverse} \] We are going to use this key relationship in reverse, i.e.
we will prove that: in our longitudinal geometry the macroscopic field (i) does not depend on what happens at the slab surfaces;  (ii) is constant in the bulk of the slab, and (iii) is a well defined linear response function of the material. This amounts to prove that $\mu_{xxxx}$ is a bulk material property, and that no surface contribution enters \equ{def} for the class of solids dealt with here.

Our key ingredient is the {\it microscopic 
electric field} $\E(\r)$ linearly induced by a unit displacement ${\bf u}_\ell$ of the $\ell$-th nucleus in the otherwise unperturbed lattice. Because of translational invariance, \[ \frac{\partial E_\beta(\r)}{\partial u_{\ell,\alpha}} = {\cal E}_{\alpha\beta}(\r - \R_\ell) \label{defx} \] and ${\cal E}_{\alpha\beta}(\r)$ is a well defined linear response of the system; it goes to zero as an inverse power of $| \r |$ and admits a multipolar expansion. We remind that the $x$-direction is not necessarily a cubic axis. The induced charge is given by $\nabla \cdot \E = 4 \pi \rho$; in a primitive lattice it is an odd function of $\r$, and its dipole is zero because of the acoustic sum rule~\cite{PCM}. Therefore the induced perturbation is octupolar to leading order.

We focus on the atomic plane whose equilibrium $x$-coordinate is $X_m = m d$, where $d$ is the interplanar distance and $m$ is an integer, and we reduce our problem to an effective one-dimensional one by averaging everything in the $yz$ plane. The averaged {\it microscopic} $\E$ field in the $x$-direction induced by a rigid displacement $u_m$ of the $m$-th atomic plane in the $x$ direction is \[ \bar{E}(x) =  u_{m}  \, \bar{\cal E}_{x x}(x - X_m) , \label{onedim}  \] where $ \bar{\cal E}_{x x}(x)$
is the average of ${\cal E}_{x x}(\r)$ in the planes normal to $x$. The $yz$-averaged induced charge is then, by Poisson equation \[ \bar{\rho}(x) = \frac{1}{4 \pi} \frac{d}{dx} \bar{E}(x) =  u_{m}  \, \frac{1}{4 \pi} \frac{d}{dx} \bar{\cal E}_{x x}(x - X_m) . \label{charge}  \] The conditions of zero induced monopole, dipole, and quadrupole lead, after integration by parts, to \bea \bar{\cal E}_{x x}(\infty) = \bar{\cal E}_{x x}(-\infty) &=& 0 , \qquad \int_{-\infty}^\infty d x \;  \bar{\cal E}_{x x}(x) = 0 , \nn \int_{-\infty}^\infty d x \; x\bar{\cal E}_{x x}(x)&=& 0  \label{f1} , \eea
while the second moment of $\bar{\cal E}$ is essentially the octupole:
\[ \int_{-\infty}^\infty d x \; x^2 \bar{\cal E}_{x x}(x) = - \frac{4 \pi}{3} {\cal Q}^{(3)}_{xxxx} \label{f2} , \] where ${\cal Q}^{(3)}_{xxxx}$ is the third moment of the induced charge. Actually, {\it all} moments of $\bar{\cal E}_{x x}(x)$ are convergent integrals, which proves that one-dimensional electrostatics is short range: $\bar{\cal E}_{x x}(x) \rightarrow 0$ faster than any polynomial for $x \rightarrow \pm \infty$.

If our slab is subject to a constant strain gradient along $x$, the displacement
of the $m$-th plane can be written as  $u_m = \eta d m^2 / 2$. In fact the local strain at plane $m$, defined as \[ \epsilon_{xx}(X_m) = \frac{u_{m+1} - u_{m-1}}{2d} , \] is thus $\epsilon_{xx}(X_m) = \eta m$ ($\eta$ dimensionless constant). The induced microscopic field, averaged over $yz$, is after \equ{onedim}: \[ \bar{E}(x) =  \frac{\eta d}{2} \sum_{m \in {\rm slab}} m^2  \, \bar{\cal E}_{x x}(x - X_m) . \label{onedim2}  \] Here we have assumed that the terms with $X_m$ close to the slab edge are identical to those in the bulk; actually, they are different, but the surface effects (due to a  nonextensive set of $m$'s) cannot propagate deep in the bulk. Owing to the short range of the perturbation within our chosen geometry, extrinsic (i.e. surface) effects are ruled out. Actually we can replace the finite sum with the infinite one, over all $m$ from $- \infty$ to $\infty$.:
\[ \bar{E}(x) =  \frac{\eta d}{2} \sum_{m =- \infty}^\infty m^2  \, \bar{\cal E}_{x x}(x - X_m) .\label{onedim3}  \]
We stress that the logics leading to our formalism for the infinite solid---via slabs of growing thicknesses---amounts to performing the thermodynamic limit in two steps: first along $yz$ and then along $x$. This guarantees that whenever the $u_m$'s in \equ{onedim} lead to a constant {\it macroscopic} field $\E$, the corresponding macroscopic polarization obeys  $\E = -4\pi\P$. 

So far we have arrived at the {\it microscopic} field induced by a uniaxial flexoelectric distortion, \equ{onedim3},
 which is a wildly oscillating nonperiodic function of $x$.
At this point we need to address the  {\it macroscopic average} of $\bar{E}(x)$, 
in order to prove that it is indeed a constant everywhere in the sample.
According to textbooks~\cite{Jackson} the macroscopic field is  the convolution  \[ \ev{E(x)} = \int_{-\infty}^\infty dx' \; w(x-x')\bar{ E}(x')  \equiv (w * \bar{E}) (x) , \label{macro} \] where $w(x)$ is a real function, nonzero in some neighborhood of $x=0$, and normalized to unity; we are going to show that the function $\bar{E}(x)$ in \equ{onedim3} becomes a constant after convolution with a suitable $w(x)$.
We start with the standard window convolution function \[ w_1(x) = \frac{1}{d} \vartheta( d/2 - |x|), \] where $\vartheta$ is the step function. Obviously $w_1 * f$ extracts the average from any function $f(x)$ periodic of period $d$. Next we define the family of convolution functions \[  w_2(x) = (w_1 * w_1)(x), \dots  w_n(x) = (w_1 * w_{n-1})(x) . \] Suppose  that $f(x)$ is a polynomial of degree $n$
times a periodic function (of period $d$) {\it whose average is zero}: it is easy to show that $(w_{n+1} * f)(x)$ vanishes in this case.

We write the microscopic field of \equ{onedim3} identically as
 \bea \bar{E}(x) &=& \frac{\eta }{2 d} \sum_m (x- md)^2   \, \bar{\cal E}_{x x}(x - md) \nn &-& \frac{\eta x}{d} \sum_m (x - md)   \, \bar{\cal E}_{x x}(x - md) \nn &+& \frac{\eta x^2 }{2 d} \sum_m    \, \bar{\cal E}_{x x}(x - md) . \label{flex2} \eea If we now choose $w_3(x)$, as defined above, to perform the macroscopic average, it is easy to see that both the second and third lines of \equ{flex2} yield a vanishing result. The term in the first line is a periodic function, whose average is obviously constant
\[ E_x = \ev{E(x)} = \frac{\eta}{2 d^2} \int_{-\infty}^\infty dx \; x^2 \bar{\cal E}_{xx}(x)  = - \frac{4 \pi \eta}{6 d^2} {\cal Q}^{(3)}_{xxxx}, \] where \equ{f2} has been used. Since our dimensionless $\eta$ equals $d \, \partial \epsilon_{xx}/\partial x$, \equ{reverse} yields \[ \mu_{xxxx} = \frac{\varepsilon_\infty {\cal Q}^{(3)}_{xxxx}}{6 d} , \label{central} \] which is the central result of this work, proving that the uniaxial flexoelectric coefficient can be expressed in terms of bulk linear response quantities.

Using the same path as above, it is immediate to verify that {\it constant} strain---i.e. $u_m = \epsilon_{xx} X_m$---induces a vanishing macroscopic field and a vanishing polarization, as it must be: a primitive lattice is nonpiezoelectric.

In order to confirm our central result, we consider next a long-wavelength longitudinal phonon, generalizing  Martin's approach~\cite{Martin72}, originally devised for  piezoelectricity, to the flexoelectric case. For a purely longitudinal phonon along $x$ of amplitude $u$ and wave vector $k$ the displacements of the atomic planes are $u_m = u \, {\rm e}^{i k X_m}$, hence the induced microscopic electric field, after $yz$ average, is \bea \bar{E}(x) &=& u \sum_{m =- \infty}^\infty   {\rm e}^{i k X_m} \, \bar{\cal E}_{x x}(x - X_m) \nn &=&  u  \, {\rm e}^{i k x} \sum_{m =- \infty}^\infty   {\rm e}^{- i k (x - X_m)} \, \bar{\cal E}_{x x}(x - X_m) 
. \label{phonon}  \eea This is the product of a slowly varying envelope function times a periodic function. The macroscopic field and polarization at wavelength $k$ are therefore
 \bea \langle E_k \rangle  &=& u \, \langle \sum_{m =- \infty}^\infty   {\rm e}^{- i k (x - X_m)} \, \bar{\cal E}_{x x}(x - X_m)\, \rangle \nn  \langle P_k \rangle  &=& - \frac{\langle E_k \rangle}{4 \pi} = - \frac{u}{4 \pi d} \int_{- \infty}^\infty  dx \;  {\rm e}^{- i k x} \, \bar{\cal E}_{x x}(x)
, \label{phonon2}  \eea  Its lowest-order $k$ expansion yields \bea \langle P_k \rangle   &\simeq& \frac{u}{4 \pi d} 
\frac{k^2}{2} \int_{- \infty}^\infty  dx  \,x^2 \bar{\cal E}_{x x}(x)
 \nn &=& - \frac{u k^2}{6 d} {\cal Q}^{(3)}_{xxxx} 
, \label{phonon3}  \eea where \equ{f2} has been used; terms of order zero and one in $k$ vanish owing to \equ{f1}. At this point we remind that the macroscopic strain induced by a long-wavelength acoustic phonon is~\cite{Martin72} $\epsilon_{xx} = i u k$, hence its gradient is $i k \, \epsilon_{xx} = -u k^2$.  \equ{phonon3} becomes \[ \langle P_k \rangle =
\frac{{\cal Q}^{(3)}_{xxxx}}{6d} \frac{\partial \epsilon_{xx}}{\partial x} . \] Comparison with our central result, \equ{central}, proves that the flexoelectric polarization induced by a long wavelength phonon and the one induced by a uniform strain gradient are the same. The trivial $\varepsilon_\infty$ factor simply accounts for the fact that our primary $\mu$ definition is transverse and not longitudinal---see \eqs{uni}{reverse}.

The present theory is strongly inspired by Martin's theory of piezoelectricity~\cite{Martin72}, although microscopic induced fields are addressed here instead of induced charges. In a piezoelectric crystal the leading multipoles are dipoles and quadrupoles: according to Martin these uniquely determine the piezoelectric tensor. In the simple class of crystals considered here dipoles and quadrupoles are both zero: we have shown that octupoles uniquely determine the flexoelectric tensor. In this sense, flexoelectricity can be regarded as the next higher order analogue of piezoelectricity.

The above considerations also hint at how the present theory can be extended to real materials of interest, for instance to cubic perovskites~\cite{Zubko07,Hong10}. In such crystal structure our basic linear-response ingredients, \equ{defx}, acquire a sublattice index, i.e. ${\cal E}_{\alpha\beta}(\r) \rightarrow {\cal E}_{s,\alpha\beta}(\r)$; the dipoles no longer vanish but their sum over $s$ does, owing to the acoustic sum rule~\cite{PCM}; the quadrupoles vanish since each nuclear site is centrosymmetric. The above derivation can therefore be extended to this case, taking into account the fact that the dipoles are coupled to the internal strain and yield a nonvanishing contribution, similarly to what done by Martin in the case of piezoelectricity~\cite{Martin72,nota2}. In a nonprimitive lattice a related qualitative feature also occurs: $\varepsilon_0 \neq \varepsilon_\infty$. According to general macroscopic considerations~\cite{Landau,rap_a30} the longitudinal and transverse polarizations differ by a factor $\varepsilon_0$, not  $\varepsilon_\infty$~\cite{nota}. This implies that the perturbed nuclear coordinates in the longitudinal and transverse geometries are {\it different}, at variance with the elemental case sketched in Fig.~\ref{fig:slabs}; this difference amounts nonetheless to an internal strain, whose effect can be tamed (once more, in the same way as in piezoelectricity~\cite{nota2}).

The present theory is also indebted to the theory of ``absolute deformation potentials''~\cite{rap61}, which addresses the electrostatic potential lineup between two differently strained regions of the same solid. Suppose we have a ``homojunction'' normal to $x$, i.e. a region on the left where the strain is zero and one on the right where the strain is $\epsilon_{xx}$. The two regions are joined by a slab where the strain is graded; $\epsilon_{xx}$ coincides with the integrated strain gradient across the slab. In the transition slab the flexoelectric response is nonzero, and a depolarizing field is present (given the longitudinal geometry); its integrated value  coincides with the potential lineup. Not surprisingly, the theory of Ref.~\cite{rap61} is based on induced octupoles as the present one is.

Last but not least, we strongly disagree with the treatment of flexoelectricity
provided by Tagantsev in Ref.~\cite{Tagantsev86}, and subsequently adopted by other authors (e.g. Ref.~\cite{Maranganti09}). This treatment is based on the rigid-ion model throughout; since a primitive lattice has zero ionic charges, the model is inadequate to address  flexoelectricity in our case study. But there is more to say: electrons behave quite differently from classical point charges, and quantization phenomena dominate macroscopic polarization, sometimes in counterintuitive ways~\cite{rap_a12}.
One of the purportedly non-bulk contributions to flexoelectricity is the  charge second moment per unit volume of the unperturbed macroscopic sample (symbol ``$I$'' in Refs~\cite{Tagantsev86,Maranganti09}). For a centrosymmetric insulating material, whose surfaces are also insulating, this quantity vanishes owing to the theorem of quantization of the surface charge~\cite{surface}; the theorem is generally violated by a system of classical point charges. Other flaws of Ref.~\cite{Tagantsev86} are the presence of conditionally convergent sums, without any prescription about their thermodynamic limit, and the lack of any relationship between shape and macroscopic field in the polarized sample.

In conclusion, we have achieved a first basic step towards a bulk theory of flexoelectricity. We have shown that---in the simplest class of dielectrics at least---there is no surface contribution to flexoelectricity, which is a therefore a pure bulk effect. Furthermore the bulk flexoelectric responses for either long-wavelength phonons or uniform strain gradient are the same; they are expressed in terms of quantities which are manifestly surface-independent.

Work supported by the ONR grant  N00014-07-1-1095.


\begin{thebibliography}{10}
\bibitem{nota2}
{ Martin's theory of piezoelectricity can be equivalently rederived using our
  ${\cal E}_{s,\alpha\beta}(\r)$ and the same logic as here}.
\bibitem{nota}
{ There is one notable exception: the polarization induced by a zone-center TO
  phonon. In this case the longitudinal and transverse polarizations differ by
  a factor $\varepsilon_\infty$}.
\end{thebibliography}

\end{document}